\documentclass[aps,prx,twocolumn,superscriptaddress,nofootinbib]{revtex4-2}
\usepackage[utf8]{inputenc}

\usepackage[titletoc,toc,title,page]{appendix}

\usepackage{csquotes,graphicx}
\usepackage[english]{babel} 

\usepackage{microtype} 

\usepackage{bm} 

\usepackage{dsfont} 
\usepackage{amsmath,amssymb,amsthm,thmtools}
\usepackage{mathtools}
\usepackage{cases}
\usepackage{calc,palatino}
\usepackage{mathrsfs} 

\usepackage{dcolumn}
\usepackage{bm}
\usepackage{xcolor}  
\usepackage{soul}
\usepackage{float}
\usepackage{physics}
\usepackage{amsfonts}
\usepackage{comment}
\usepackage{nameref}
\usepackage[colorlinks=true]{hyperref}
\usepackage[nameinlink]{cleveref}
\crefname{appsec}{Appendix}{Appendices}
\crefname{box}{Box}{Box}
\hypersetup{
  colorlinks   = true, 
  urlcolor     = green!80!black, 
  linkcolor    = blue, 
  citecolor    = red!80!black 
}

\newcommand{\pbmelo}[1]{\textcolor{RoyalBlue}{\fbox{PBM} #1}}
\newcommand{\simone}[1]{\textcolor{purple}{\fbox{SA} #1}}

\usepackage[dvipsnames,usenames]{xcolor}

\newcommand{\mauro}[1]{\textcolor{NavyBlue}{#1}}

\newcommand{\parTitle}[1]{\textit{#1 ---}}

\usepackage{ulem}

\begin{document}

\title{Non-equilibrium thermodynamics of collapse models in the strongly non-Gaussian regime}

\author{Pedro B.~Melo$^{\star}$}
\email{pedrobmelo@aluno.puc-rio.br}
\affiliation{Universit\`{a} degli Studi di Palermo, Dipartimento di Fisica e Chimica - Emilio Segr\`{e}, via Archirafi 36, I-90123 Palermo, Italy}
\affiliation{Departamento de F\'isica, PUC-Rio, 22452-970, Rio de Janeiro RJ, Brazil}

\author{Pedro V.~Paraguass\'u$^{\star}$}
\email{paraguassu@esp.puc-rio.br}
\affiliation{Departamento de F\'isica, PUC-Rio, 22452-970, Rio de Janeiro RJ, Brazil}

\author{Simone Artini$^{\star}$}
\email{simone.artini@unipa.it}
\affiliation{Universit\`{a} degli Studi di Palermo, Dipartimento di Fisica e Chimica - Emilio Segr\`{e}, via Archirafi 36, I-90123 Palermo, Italy}

\author{Gabriele Lo Monaco}
\affiliation{Universit\`{a} degli Studi di Palermo, Dipartimento di Fisica e Chimica - Emilio Segr\`{e}, via Archirafi 36, I-90123 Palermo, Italy}

\author{Sandro Donadi}
\affiliation{Universit\`{a} degli Studi di Palermo, Dipartimento di Fisica e Chimica - Emilio Segr\`{e}, via Archirafi 36, I-90123 Palermo, Italy}

\author{Mauro Paternostro}
\affiliation{Universit\`{a} degli Studi di Palermo, Dipartimento di Fisica e Chimica - Emilio Segr\`{e}, via Archirafi 36, I-90123 Palermo, Italy}
\affiliation{Centre for Quantum Materials and Technologies, School of Mathematics and Physics, Queen's University Belfast, BT7 1NN, United Kingdom}

\begin{abstract}
Standard objective collapse models offer a unified approach to the quantum measurement problem but predict an unphysical, indefinite increase in the energy of the system. The dissipative Diósi-Penrose (dDP) model resolves this heating issue by introducing a linear friction mechanism. However, this modification induces complex, non-Gaussian phase-space dynamics. We rigorously establish the thermodynamic consistency of this friction mechanism - extended to the CSL model - across both weakly and strongly non-Gaussian regimes. Using the Wigner phase-space formalism, we go significantly beyond the quadratic approximation and, to bypass the failure of perturbative methods under strong dissipation, introduce a novel exact pseudo-spectral simulation approach. Our analysis reveals that the system subjected to the dDP mechanism does not thermalize, but rather settles into a non-equilibrium steady-state (NESS) where the asymptotic non-Gaussianity scales as the third power of the dissipation parameter $\beta$. By evaluating the Wigner entropy production, we confirm the thermodynamic validity of the model and demonstrate that highly sensitive information--theoretic quantities require exact numerical methods to accurately capture the key non-Gaussian tails of the distribution.
\end{abstract}

\maketitle
\def\thefootnote{$\star$}\footnotetext{These authors contributed equally to this work}\def\thefootnote{\arabic{footnote}}

\section{Introduction}
While the role of measurement lies at the heart of the quantum-to-classical transition and  macro-objectification -- \textit{i.e.} the suppression of linear superpositions of macroscopic objects -- how exactly this process takes place is still subject to debate. Collapse models \cite{ghirardi1986unified,ghirardi1990markov,pearle1989combining,diosi1987universal,diosi1989models,penrose1996gravity} offer a potential resolution 
by adding a non-linear and stochastic mechanism to the otherwise linear  Schr\"odinger equation~\cite{carlesso2022present, Carlesso2025}. Among several formulations, some of the most celebrated versions of collapse models are the Continous Spontaneous Localization (CSL) model \cite{ghirardi1990markov,bassi2003dynamical}  and the Di\'osi-Penrose (DP) one \cite{diosi1987universal,diosi1989models}. 

Collapse models are specifically designed to describe the suppression of coherent superpositions of macroscopic objects, while leaving microscopic systems essentially unperturbed \cite{bassi2003dynamical, carlesso2022present}. Their phenomenological nature implies that the thermodynamic consistency of both the CSL and the DP model is not guaranteed and, in fact, they predict an indefinite, and thus non-physical,  increase in the mean energy of the system, associated with a constant heating power.  
To overcome this issue, a dissipative version of both the CSL \cite{smirne2015dissipative} and the DP \cite{bahrami2014role,di2023linear} models has been introduced. 

Physical legitimacy is not guaranteed by requiring that such  dynamics have thermal states at finite temperature as asymptotic states. In particular, any phenomenological model describing the quantum-to-classical transition must adhere to the Second Law of Thermodynamics and the sole addition of a dissipative mechanism is not sufficient to ensure positive entropy production. The study of the thermodynamics of a system described by a collapse models is a non-trivial task, as we need to track its  inherently non-equilibrium dynamics. 
Remarkably, attempts at the characterization of the thermodynamics of the CSL and DP model have been made, for the first time, recently~\cite{artini2023characterizing,artini2025nonequilibrium, te2021master, Campbell_2026}. A methodological route to the study of the thermodynamic consistency of dissipative collapse models passes through the assessment of the positivity of the Wigner entropy production rate \cite{landi2021irreversible,Santos2017,santos2018irreversibility}, as it naturally extends the classical stochastic thermodynamics framework in many relevant scenarios \cite{artini2025non}. Ref.~\cite{artini2025nonequilibrium} focused primarily on a specific regime of the dissipative DP model in which Gaussianity is preserved by the dynamics, allowing the phase-space formalism to be exploited to its full extent. However, the frictional extensions of CLS ad DP models present a much richer dynamics once one enters the non-Gaussian regime.  

In this paper, we examine the CSL counterpart of the dissipative DP model (dDP) with linear-friction as defined in Ref. \cite{di2023linear} in full details, including regimes where the non-Gaussianity of the state becomes strong and non-negligible. Given the high complexity of the model, here we propose a novel algorithm that allows to resolve the non-Gaussian dynamics with high precision even for strong non-Gaussianity and long times. At technical level, the proposed algorithm leverages both a Gram-Charlier (GC) expansion and a pseudospectral algorithm based on exponential time differencing to simulate the non-Gaussian evolution of the Wigner function of a system subject to this collapse mechanism. 

We find that the first non-Gaussian contribution affects the moments of the distribution only starting from the fourth order, causing a bigger kurtosis and thus fatter tails (lepto--kurtosis) with respect to a reference Gaussian distribution. This apparently simple observation has deep consequences, since it allows the use the GC expansion for small values of the friction parameter $\beta$ [introduced in \cref{eq: lindblad op dDP}], which is related to non-Gaussianity. In addition, the pseudospectral algorithm is used to broaden the range of applicability of our method to cases where larger deviations from Gaussianity are present. Our analysis explicitly quantify the non-Gaussianity $\delta$ [cf. \cref{eq: non gauss meas}] stemming from the dynamics, finding a polynomial relation with the dissipation parameter $\delta \propto \beta^3$, and the non-equilibrium entropy rate produced in the evolution. In particular, we show how the dDP model achieves thermal equilibrium only in the small $\beta$ limit, whereas for finite values the system reaches a highly non-equilibrium steady-state distribution. Ultimately, our findings establish the thermodynamic consistency of the non-Gaussian dDP model, highlighting the crucial role of phase-space tails in macroscopic quantum objectification.

The paper is organized as follows: in \cref{Sec:2}, we introduce the necessary background about collapse models, emphasizing dissipative variants and their phase-space representation via quantum Fokker-Planck equation\footnote{In fact, the master equation adopted here has terms that go further from Fokker-Planck approximation, but we term it this way to make it easier to understand.}. In \cref{sec: thermo stuff}, we outline the non-equilibrium thermodynamic framework used to assess the consistency of the non-Gaussian dynamics studied in this work. \Cref{sec: moments} presents the system of differential equations used to derive analytical solutions for the statistical moments. In \cref{sec: gram_charlier}, we introduce our first approach to computing Wigner function evolution using the Gram-Charlier expansion, which approximates the distribution in regimes near Gaussianity. \Cref{sec: simulation} details a pseudo-spectral simulation method that is not limited to weak non-Gaussianity. Finally, in \cref{sec: conclusions}, we discuss our results and provide concluding remarks.

\section{Collapse models in phase-space \label{Sec:2}}
The unitary linear evolution that lies at the core of standard QM imposes severe constraints on the dynamics. Collapse models propose to restore the consistency with experienced macroscopic objectivity giving up the linearity in the Schr\"odinger equation.
The modified Schr\"odinger equation governing the dynamics of the standard formulation of the collapse models is 

\begin{equation}\label{eq: collapse Schrodinger}
\begin{aligned}
& d|\psi_t\rangle =  \left[ -\frac{i}{\hbar} \hat{H}dt + \int d^3x \left( \hat{M}(x) - \langle \hat{M}(x) \rangle_t \right) dW_t(x) \right. \\
& \left. - \frac{1}{2} \int d^3x d^3y \mathcal{D}(x-y) \prod_{q=x,y} \left( \hat{M}(q) - \langle \hat{M}(q) \rangle_t \right) dt \right] |\psi_t\rangle \,,
\end{aligned}
\end{equation}
where $\hat{H}$ is the standard system Hamiltonian, $\hat{M}(x)$ is the mass density operator that ensures that the effect of the collapse is amplified as the system's size increases, and $dW_t(x)$ represents the Brownian noise driving the collapse. The term $\langle \hat{M}(q) \rangle_t = \langle \psi_t | \hat{M}(q) | \psi_t \rangle$ accounts for the nonlinear contribution required to guarantee that the dynamics genuinely collapses the state vector. The function $\mathcal{D}(x-y)$ differentiate the CSL and the DP models and it characterizes the spatial resolution of the collapse. The CSL model assumes a Gaussian correlator defined as $\mathcal{D}_{\text{CSL}}(x-y) = \frac{\lambda}{m_0^2} \exp\left( -\frac{|x-y|^2}{4r_C^2} \right)$, where $\lambda$ is the collapse rate, which determines the strength of the collapse, and $r_C$ is typical localization length. The DP model relates the collapse to gravity using a correlator proportional to the Newtonian potential $\mathcal{D}_{\text{DP}}(x-y) = \frac{G}{\hbar} \frac{1}{|x-y|}$, where $G$ is the gravitational constant. Moreover, in the DP model, the mass density $\hat M(x)=\sum_i m_i \delta(\hat q_i-x)$ is replaced by a smeared out mass density $\hat M(x)=\sum_i m_i g(\hat q_i-x)$ with $g(z)=\,(2\pi R_0^2)^{-3/2}
e^{-z^2/(2R_0^2)}$, required to avoid divergencies the model encounters when applied to point particles.

Starting from \cref{eq: collapse Schrodinger}, one can study the evolution of the statistical operator $\hat{\rho}_{t} = \mathbb{E}[|\psi_{t}\rangle\langle\psi_{t}|]$, representing the stochastic average with respect to the collapse noise. The equation describing the evolution of the statistical operator is of the standard Lindblad \cite{bassi2003dynamical} type
\begin{equation}
\label{eq: collapse ME}
\frac{d}{dt}\hat{\rho}_{t} = -\frac{i}{\hbar}[\hat{H}, \hat{\rho}_{t}] +\mathcal{L}[\hat \rho_t]\,,
\end{equation}
where the Lindbladian $\mathcal{L}[\hat \rho_t]$ can be written either as
\begin{equation}
   \mathcal{L}[\hat \rho_t] =  \int d^{3}x d^{3}y \mathcal{D}(x-y) [\hat{M}(x), [\hat{M}(y), \hat{\rho}_{t}]]\,,
\end{equation}
or equivalently, moving to Fourier space, as \cite{di2023linear}
\begin{equation}
\label{eq: frictionless DP}
    \mathcal{L}[\hat \rho_t]=\frac{1}{\hbar^2}\int \frac{d^3k}{(2\pi)^3}\Gamma(k)\left(\hat L_k\hat \rho_t \hat L^\dagger_k-\frac{1}{2}\left\{\hat L^\dagger_k\hat L_k,\hat \rho_t \right\}\right)\,.
\end{equation}
The two models share the same jump operators, $\hat{L}_k=me^{ik\cdot\hat{x}}$, while the function $\Gamma(k)$ is again model dependent, as the correlator. In particular
\begin{equation}
\begin{aligned}
    \Gamma_{\rm DP}(k)&=\frac{4\pi\hbar G}{k^2}e^{-k^2 R_0^2}\\ 
    \Gamma_{\rm CSL}(k)&=\frac{\gamma \hbar^2\bar{m}^2}{8m^2\pi^3}e^{-k^2 r_C^2/\hbar^2}
\end{aligned}
\end{equation}
with $\gamma = 8\pi^{\frac{3}{2}}r_C^3\lambda$ and $\bar m$ being a dimensionless mass written in units of the reference mass $m_0=1$ amu. While the underlying wavefunction evolution is nonlinear and stochastic, the dynamics of the density matrix in \cref{eq: collapse ME} is linear. This linearity is a critical feature of the theory, as it ensures the preservation of causality by forbidding superluminal signaling \cite{gisin1989stochastic}, even though the collapse itself is a non-local process.

The dissipative extensions of the DP and he CSL models introduced in \cite{di2023linear} differ from the frictionless couterpart because of the form of the the Lindblad operators:
\begin{equation} \label{eq: lindblad op dDP}
\hat L_k=me^{ik\cdot \hat x}-\frac{\hbar \beta}{8}\left\{k\cdot \hat p,e^{ik\cdot \hat x}\right\}\,,
\end{equation}
where $\beta$ is the parameter that regulates the strength of the dissipation.

In order to study the dynamics of a system subject to the collapse dynamics, we move to the phase-space description. The Lindblad equation becomes the equation of motion for the Wigner function os the state. In the frictionless scenario, both the collapse models yields the same differential equation of Fokker--Planck type \cite{artini2023characterizing,artini2025nonequilibrium}: 
\begin{equation}\label{eq:Wigner_diss}
\partial_tW_{\hat\rho(t)}\,=\,\{W_{\hat H},W_{\hat \rho(t)}\}_\star+D\,\Delta_{(\bm p)}W_{\hat \rho(t)}\,,
\end{equation}
where $\{W_{\hat H},W_{\hat \rho(t)}\}_*$ is the Moyal bracket that encodes the unitary part of the dynamics \cite{baker1958formulation} and $D$ is the positive diffusion constant that depends on the parameters of the model: $D_{DP}=\frac{G\hbar m^2}{3\sqrt{\pi} R_0^3}$ and $D_{\rm CSL} =  \frac{\lambda}{r_c^2} \frac{m^2}{m_0^2}$. This Fokker-Planck evolution is obtained assuming a  Wigner function that is well concentrated around the origin and using the Kramers--Moyal expansion (which is just a Taylor expansion up to the second order of $W(q,p+p')$ in $p$, see \cite{artini2025nonequilibrium}).  Under this approximation, the evolution becomes Gaussian. 

It is possible to write a similar differential equation governing the evolution of the Wigner function in the dissipative case of the dDP model, again using the Kramers--Moyal expansion. One gets  \cite{artini2025nonequilibrium}: 
\begin{equation}
\label{eq: dDP Wigner}
\begin{aligned}
    \partial_t W_{\hat \rho}&=\left\{W_{\hat H},W_{\hat \rho}\right\}_\star+\nabla_{\bm p}\!\cdot\!(f\bm p W_{\hat \rho})+\partial_{p_i}\partial_{p_j}(D_{ij}W_{\hat \rho})\\
    &+\partial_{q_i}\partial_{q_j}(\tilde D_{ij}W_{\hat \rho})+ R^{ijkl}\partial_{p_i}\partial_{p_j}\partial_{q_j}\partial_{q_k}W_{\hat \rho},
\end{aligned}
\end{equation}
where Einstein's notation for the summations is understood. In the following, we will focus on the one-dimensional problem for simplicity. Notice how this equation of motion is not Gaussian even under this approximation. It must be stressed that the dissipator of the DP model is strictly valid in three dimensions, being it designed to be the Green function solution of the three-dimensional Laplace problem. In order to avoid a misleading interpretation of the result, we will focus instead on the CSL model. In one dimension, the resulting equation greatly simplifies
\begin{equation}
\label{eq: working model}
\begin{aligned}
    & \partial_t W = D_{pp} \partial_p^2 W + D_{qq}\partial_q^2 W + \left(F_1 + 2F_2\right)W+\\
	&+ \left(F_1+4F_2\right) p \partial_p W 
    +F_2 p^2 \partial_p^2 W + D_{q^2p^2} \partial_q^2 \partial_p^2 W.
\end{aligned}
\end{equation}
The coefficients $D_{ij}$ and $F_i$ are constants that depend on the physical parameters of the system: the force components $F_1$ and $F_2$ are given by
\begin{align}\label{eq:F1F2}
	F_1 &= \frac{\beta m \hbar^2 \gamma \sqrt{\pi}}{4\pi R_0^3} - \frac{3 \hbar^4 \beta^2 \gamma \sqrt{\pi}}{32\pi R_0^5},  \\
	F_2 &= \frac{3 \hbar^4 \beta^2 \gamma \sqrt{\pi}}{64\pi R_0^5}\,,
\end{align}
while the diffusion coefficients, which characterize the stochastic effects of the environment, are
\begin{align}
	D_{pp} &= \frac{m^2 \gamma \hbar^2}{\pi} \left( \frac{\sqrt{\pi}}{2R_0^3} + \frac{3 \hbar^2 \beta \sqrt{\pi}}{8m R_0^5} + \frac{15 \hbar^4 \beta^2 \sqrt{\pi}}{128 m^2 R_0^7} \right), \\
	D_{qq} &= \frac{\hbar^4 \beta^2 \gamma \sqrt{\pi}}{32\pi R_0^3},\,	D_{q^2p^2} = \frac{3 \hbar^6 \beta^2 \gamma \sqrt{\pi}}{64\pi R_0^5}\,.
\end{align}

\section{Non-equilibrium thermodynamics}
\label{sec: thermo stuff}
We now introduce the main tools used to investigate the thermodynamics of the collapse models, on the same line as \cite{artini2023characterizing,artini2025non,artini2025nonequilibrium}. In nonequilibrium thermodynamics, it is common to consider the following formulation of the Second Law of Thermodynamics:
\begin{equation}
\label{eq:Clausius}
    \Sigma = \Delta\text{S}_{\text{sys}} +  \Phi\geq 0\,,
\end{equation}
where $\Delta\text{S}_{\text{sys}}$ is the variation of the entropy in the system, $\Phi$ represents the entropy flux between the system and the environment, and $\Sigma$ is the entropy production, that is the amount of entropy lost in the transformation due to friction, which is always non-negative due to Clausius' Theorem and zero only at equilibrium. The entropy production quantifies the degree of irreversibility of the process the system is undergoing and different formulations of this quantities are found across all length scales. It is often convenient to rewrite \cref{eq:Clausius} in a differential form defining the entropy production rate $\Pi$ and the entropy flux rate $\phi$,
\begin{equation}
    \dot{\text{S}}_{sys}= \Pi(t) - \phi(t) \,.
\end{equation}
Note that $\Pi(t) \geq 0$ is also required for consistency with the Second Law of Thermodynamics.
What is the correct definition of entropy to use in quantum thermodynamics has been object of study for a long time \cite{Adesso2013,Santos2017,landi2021irreversible}. Building on \cite{Santos2017, landi2021irreversible, artini2023characterizing, artini2025nonequilibrium, artini2025non}, we consider the Wigner entropy $S_W=-\iint \,dp\,dq\,W(q,p)\ln W(q,p)$ as the proper definition of system's entropy $S_{sys}$ to use in order to formulate a consistent quantum thermodynamical framework. For states with positive Wigner functions, this quantity coincides with the Rényi$-2$ entropy $S_2 = -\ln({\rm Tr}\hat{\rho}^2)$, up to an irrelevant constant \cite{Adesso2013}. The respective entropy production rate can be thus written as 
\begin{equation}
\label{eq: EP}
\begin{aligned}
    \Pi(t)&= -\frac{d}{dt} K(W(t)||W_{\infty})\\
    &:= -\frac{d}{dt} \int dq\, dp\, W(q,p,t)\ln\frac{W(q,p,t)}{W_{\infty}(q,p)}\,,
\end{aligned}
\end{equation}
where $W_{\infty}$ is the Wigner function of a suitable target state, not necessarily the equilibrium one and $K(p||q)$ denotes the Kulback-Leibler divergence between the distributions $p$ and $q$  \cite{cover1991information}. At this point, one may wonder why the Wigner function is to prefer to the Von Neumann analogue $\Pi_{VN}(t):=-\frac{d}{dt} S_{VN}(\rho(t)||\rho_{\infty})$. The main reason is that the latter shows a pathologic behavior, the \textit{ultracold catastrophe}, \textit{i.e.} a divergence in the entropy production rate in the limit of zero temperature baths, and thus it cannot lead to a consistent framework \cite{landi2021irreversible}. It is important to stress that \cref{eq: EP} accounts for only the entropy production rate associated to the relaxation process that the system undergoes while reaching the steady--state \cite{deffner2011nonequilibrium}. If the steady--state is a non--equilibrium one, then an extra constant entropy production rate is present due to an external driving \cite{seifert2005entropy} or a non--equilibrium environment, \textit{e.g.} systems in contact with multiple baths \cite{de2016quantum,de2023quantum}. 

Using this framework, the non-equilibrium thermodynamics of the standard collapse models and of the dDP model has been characterized in~\cite{artini2023characterizing,artini2025nonequilibrium}. The frictionless CSL and DP models, whose dynamics is described in terms of the same quantum Fokker-Planck equation (eq. \eqref{eq:Wigner_diss} above), both show an indefinite heating, as consequence of the indefinite energy increase in time. Thus, in order to have an entropy production rate that is always positive along the dynamics, one should interpret the dynamics as that of a system in contact with a bath at infinite temperature. For this reason, the system never reaches thermal equilibrium. On the other hand, in the dDP model a friction mechanism guarantees a finite asymptotic energy. Besides, the dDP model presents a much richer dynamics, as the last two terms of \cref{eq: working model} breaks the Gaussian character of the evolution. A possible way to investigate the thermodynamics of the dissipative case is to consider \cref{eq: working model} in a nearly-Gaussian regime where $\beta$ is small and all the terms of order $\beta^2$ can be neglected. In this case, the Wigner function's equation of motion is again a Klein-Kramers equation, that is guaranteed to have positive entropy production rate at all times and thermalization in the long time limit. The non-Gaussian regime instead is only partially investigated using a linear expansion of the equation of motion valid for very small times. For certain values of the parameters, a negative entropy production was found in this way in \cite{artini2025nonequilibrium}, but there were no guaranties that this negativity is physical and not due to having pushed the perturbative approach beyond its limit of validity. In the analysis carried out in the next sections, we show that indeed the violation disappears once the dynamics is studied using the methods we propose in the following sections. 

\begin{figure*}
    \centering
    \includegraphics[width=0.9\linewidth]{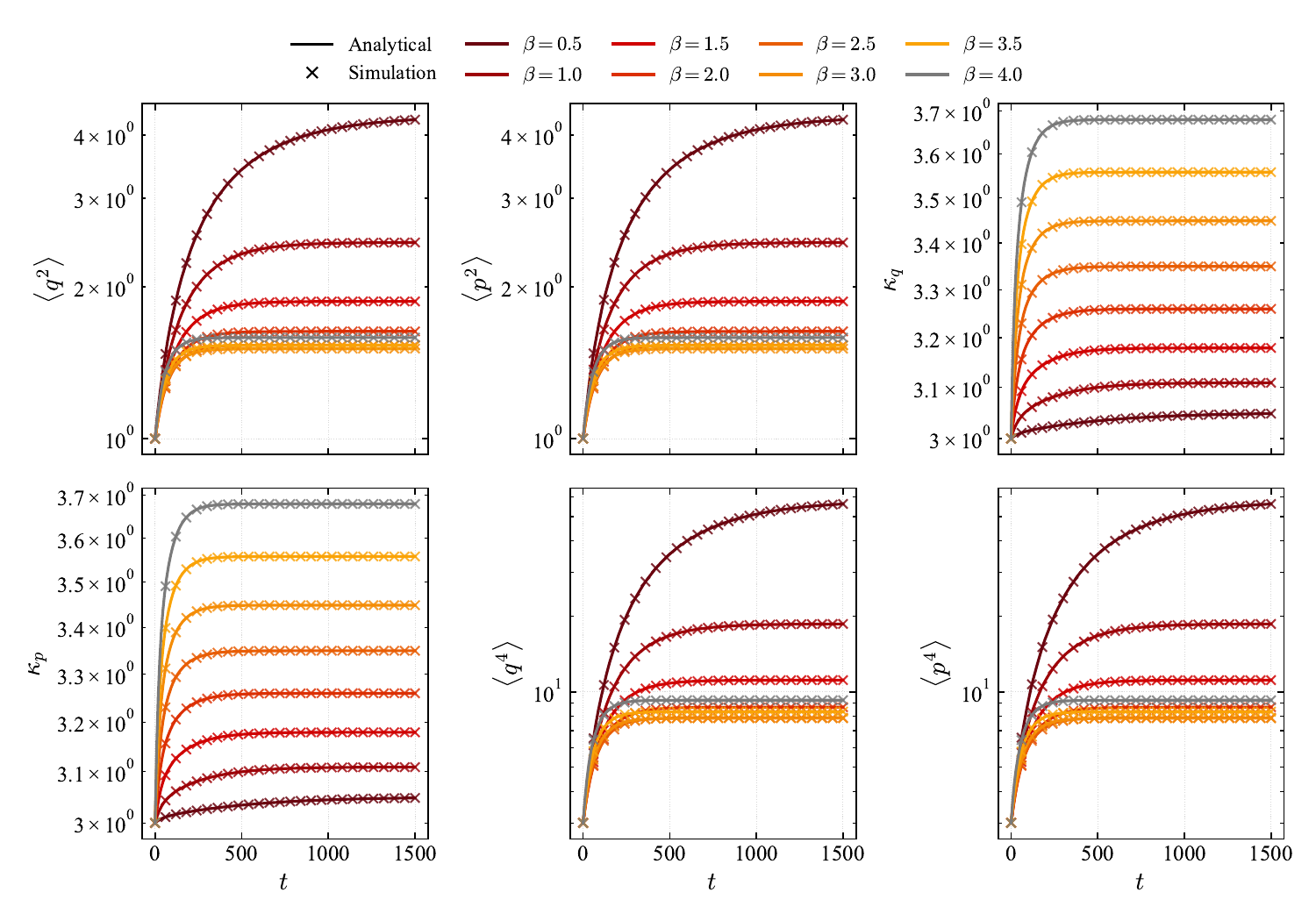}
    \caption{Time evolution of the non-zero moments for a particle in a harmonic trap governed by \cref{eq: dDP Wigner}. Solid lines represent the analytical simulation results for different values of $\beta$, while the markers (crosses) denote the numerical solutions derived from the system of ODEs. The graphs of $\kappa_{q}$ and $\kappa_{p}$ denote the kurtosis given by $\kappa_{q} \equiv \langle q^4\rangle/\langle q^2\rangle$ and $\kappa_{p} \equiv \langle p^4\rangle/\langle p^2\rangle$. The steady-state regime is reached as $t \to \infty$, where the moments $\langle q^2 \rangle$, $\langle p^2 \rangle$, and $\langle q^4 \rangle$ plateau at values determined by the competition between diffusion and dissipation.}
    \label{fig:moments_Gram_Chalier}
\end{figure*}

\section{Moments Evolution and the Onset of Non-Gaussianity} \label{sec: moments}

The study of the evolution of Wigner function beyond the Gaussian regime can be an extremely hard task. In fact, a general non-Gaussian state is determined in principle by all its momenta. In order to get some insight about the evolution of the state of the system, we start by analyzing the time evolution of the moments of the Wigner quasiprobability distribution. Let us remember that the moments are defined as:
\begin{equation}\label{eq:def_momenta}
\langle q^n p^m\rangle\equiv \int q^n p^m\, W(t)\,dq\,dp =\Tr[\rho(t)\,{\rm sym}(\hat q^n\hat p^m)]\,,
\end{equation}
where $\hat q$ and $\hat p$ denotes the position and momentum operator and the symmetrization of their product ${\rm sym}(\hat q^n \hat p^m)$. For instance,
${\rm sym}(\hat q\,\hat p)=\frac{1}{2}(\hat q\,\hat p+\hat p \hat q)$. Inserting the definition \cref{eq:def_momenta} in \cref{eq: working model}, it is straightforward to obtain the equation of motions for any moment by simple integration by part. For sake of clarity, in the following we write down explicitly the equations for the lower moments, from the first to the fourth, because they will be the most relevant for our discussion.

\parTitle{First moments} In this case, we have a set of two equations only depending on the first moment only
\begin{align}\label{eq: 1 moments}
    \frac{d}{dt} \langle q \rangle &= \frac{1}{m}\langle p \rangle , \quad    \frac{d}{dt} \langle p \rangle = -F_1 \langle p \rangle -k \langle q\rangle .
\end{align}
This implies that for an initial state centered at the origin $\langle q \rangle_0 = \langle p \rangle_0 = 0$ the time derivatives vanish, and the first moments remain zero for all time and the center of the distribution remains stationary at the origin. In the following, we will assume vanishing first moments for simplicity.

\parTitle{Second moments} Under the assumption of vanishing first moments, the second moments essentially coincide with the covariance matrix. The evolution of the covariance is thus dictated by the following closed set of three equations:
\begin{align}\label{eq: 2 moments}
    \frac{d}{dt} \langle q^2 \rangle &= 2D_{qq}+\frac{2}{m}\langle q p \rangle, \\
    \frac{d}{dt} \langle p^2 \rangle &= 2D_{pp} - 2(F_1-F_2) \langle p^2 \rangle - 2 k \langle q p\rangle, \\
    \frac{d}{dt} \langle q p \rangle &= - F_1 \langle q p \rangle + \frac{1}{m}\langle p^2 \rangle - k \langle q^2 \rangle.\label{eq:qp}
\end{align}
Equations (\ref{eq: 1 moments}) and (\ref{eq: 2 moments}) are functionally equivalent, provided $F_1>F_2$, to the ones obtained in the CPTP version of the Caldeira--Leggett model of Quantum Brownian motion \cite{breuer2002theory}. Therefore, up to this order eq.(\ref{eq: dDP Wigner}) resembles a Gaussian evolution: any initial Gaussian state would remain Gaussian if the dynamics were truncated at this level. Notice that if $F_1<F_2$, the second of equations (\ref{eq: 2 moments}) predicts a diverging momentum and thus a diverging energy of the particle. This is consistent with the discussion in \cref{sec:free particle} where it is shown that one must have $\beta<{16m R_0^2}/{(9\hbar^2)}$ to get finite second moments in the steady--state distribution, which is equivalent to asking $F_1>F_2$.

\parTitle{Third moments} When looking at the third moments, we start observing couplings between moments of different order. In particular, the dynamics of the third moments depends on the first ones, that act as sources:
\begin{align}
    \frac{d}{dt} \langle q^3 \rangle &= 6 D_{qq} \langle q \rangle + 3\frac{\langle q^2p \rangle}{m}, \\
    \frac{d}{dt} \langle p^3 \rangle &= 6 D_{pp} \langle p \rangle - (3F_1-6F_2) \langle p^3 \rangle, \\
    \frac{d}{dt} \langle p^2 q \rangle &= 2 D_{pp} \langle q \rangle - 2(F_1-F_2) \langle p^2 q \rangle + \frac{\langle p^3 \rangle }{m}, \\
    \frac{d}{dt} \langle p q^2 \rangle &= 2 D_{qq} \langle p \rangle -F_1 \langle q^2 p \rangle + \frac{2}{m}\langle p^2 q \rangle \,.
\end{align}
 Given our initial condition where all the odd moments are vanishing by symmetry, the third moments remain zero for all time as well. Writing down the equations of motion for any odd moment, one can observe that they only contains odd moments at any order and the distribution does not develop any asymmetry. Therefore, the first deviations from Gaussianity are expected to appear in the behavior of the fourth moments.

\parTitle{Fourth moments} The fourth moments are actually the most relevant to our discussion, because we finally observe deviations from Gaussianity. The evolution of the fourth moments is dictated by the set of five equations, also coupling to the covariances acting as sources as described in Appendix \ref{app:A}.

In order to determine the terms that break the Gaussian character of the dynamics, one has to consider how equations \cref{eq: 4 moments 1} to \cref{eq: 4 moments 5} deviate from those stemming from a linear Langevin dynamics associated with equations \cref{eq: 2 moments}. In particular, owing to Wick's theorem, a Gaussian dynamics has to preserve the following relations between fourth and second moments
\begin{equation}
    \begin{aligned}
\langle q^4 \rangle &= 3 \langle q^2 \rangle^2, \quad \langle p^4 \rangle = 3 \langle p^2 \rangle^2 \\
\langle q^3 p \rangle &= 3 \langle q^2 \rangle \langle qp \rangle, \quad
 \langle q p^3 \rangle = 3 \langle p^2 \rangle \langle qp \rangle \\
\langle q^2 p^2 \rangle &= \langle q^2 \rangle \langle p^2 \rangle + 2 \langle qp \rangle^2.
\end{aligned}
\end{equation}
A direct check built when considering second and fourth moments show that these relations do not hold for our dynamics. 
Considering for example $\langle p^4 \rangle$, from Wick's theorem to hold one should have $\frac{d}{dt}\langle p^4\rangle = 12D_{pp}\langle p^2 \rangle -12(F_1-F_2)\langle p^2\rangle^2 -12k\langle pq \rangle\langle p^2 \rangle$. Following again the Gaussian assumption, we can write $\langle p^4\rangle = 3\langle p^2\rangle^2$ and $\langle p^3 q\rangle=3\langle p^2\rangle\langle pq\rangle$. Then, comparing this with the equation in \cref{eq: 4 moments 4} one finds an extra $24 F_2$ that breaks this condition.


These equations confirm that the non-Gaussianity due to the friction terms in the CSL model affect mainly the fourth moments. In particular, we will see that this deviation leads to a higher kurtosis--with respect to a Gaussian--and thus fatter tails. See~\cref{fig:moments_Gram_Chalier} for an illustration of the evolution of the moments.

\begin{figure*}
\includegraphics[width=0.7\textwidth]{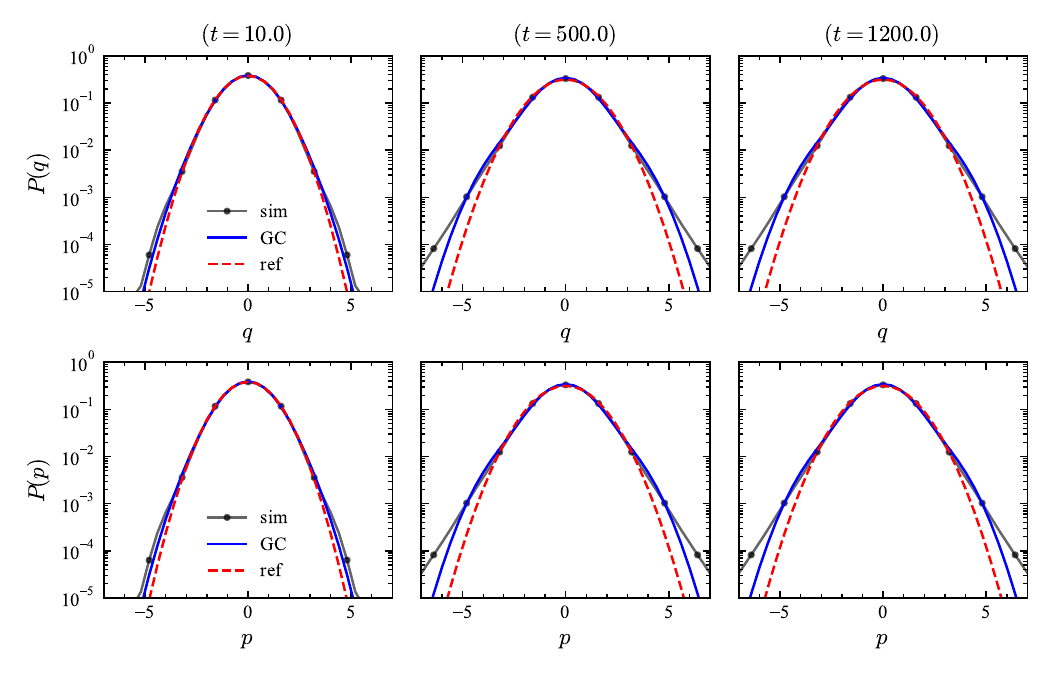}\\
    \caption{Temporal evolution of the marginal PDFs. The plots display projections (upper) onto the $q$ and (lower) $p$ axes for parameters $m=\omega = 1.0$, $R_0 = 3.0$, $\gamma=1.0$, $\beta = 4.0$ and $\hbar=k_B=1.0$  in Log scale, computed using the Gram-Charlier approximation  (solid line) and the reference Guassian marginals (dashed lines). Snapshots are taken at $t = 10$, $t = 500$, and $t = 1200$. At early times ($t = 10$), the function $W(q,p)$ closely resembles a Gaussian distribution due to the initial conditions. As the system evolves, deviations from Gaussianity emerge, driven by non-zero fourth-order moments. This non-Gaussian behavior manifests as a modification of the distribution tails and an increase in the peak height of the marginal distributions. Throughout this parameter regime, the distribution remains sufficiently close to the Gaussian reference, confirming the validity of the approximation method.}
    
\label{fig:Gram_Chalier_Wigner_and_Marginals}
\end{figure*}

\section{Approximation of the Wigner quasiprobability distribution for weak non-Gaussianity \label{sec: gram_charlier}}
\subsection{Gram-Charlier expansion}
To address the non-Gaussianity introduced by the linear friction term, we employ a Gram-Charlier expansion. This method allows us to construct an approximate non-Gaussian Wigner function from the reference Gaussian distribution by incorporating moments higher than the second order. Because we observed that all the odd moments are identically vanishing at any time, the first non-trivial correction to Gaussianity can be incorporated as follows \cite{Barton1952,Blinnikov1998}:
\begin{equation}
    W(q,p,t) \approx W_0(q,p,t)[1 + \Phi_4(q,p,t)],
\end{equation}
where we introduced a Gaussian reference $W_0(q,p,t)$: 
\begin{equation}
   W_0(q,p,t) = \frac{\exp\left\{-\textbf{x}^T\Theta^{-1}(q,p,t)\textbf{x}\right\}}{2\pi\sqrt{\det[\Theta(q,p,t)]}}.
\end{equation}
$\textbf{x}=[q,p]$ and $\Theta(q,p,t)$ denotes the covariance matrix in terms of the second-order moments ($\langle q^n p^m \rangle$ where $n + m=2$) as defined in \cref{eq:def_momenta}, written as
\begin{equation}
    \Theta(q,p,t) = \mqty(\sigma_q^2(t) & \sigma_{qp}^2(t) \\ \sigma_{qp}^2(t) & \sigma_p^2(t)),
\end{equation}
with $\sigma^2_{q} = \langle q^2 \rangle - \langle q \rangle^2$, and similarly for the other components. The term $\Phi_4(q,p,t)$ encapsulates the deviations from the Gaussian regime caused by the non--Gaussian fourth--moments, and it is given by
\begin{align}
    \Phi_4(q,p,t) &= \frac{\kappa_{q^4}}{24} H_4(q) + \frac{\kappa_{q^3p}}{6}H_3(q)H_1(p) \nonumber\\ 
    &\quad + \frac{\kappa_{q^2p^2}}{4}H_2(q)H_2(p) + \frac{\kappa_{p^3q}}{6}H_3(p)H_1(q) \nonumber\\
    &\quad + \frac{\kappa_{p^4}}{24} H_4(p),
\end{align}
where $\kappa_{q^i p^j} = \langle (q - \langle q \rangle)^i (p - \langle p \rangle)^j \rangle$ represents the cumulants, and $H_i(x)$ are the Hermite polynomials:
\begin{align}
    H_1(x) &= x,\\
    H_2(x) &= x^2 - 1,\\
    H_3(x) &= x^3 - 3x,\\
    H_4(x) &= x^4 - 6x^3 + 3. 
\end{align}

In order to improve the efficiency of the Gram-Charlier method, it is convenient to introduce a new rotated and translated set of coordinates $(x,y)$ with zero mean (when needed) and identity covariance matrix. In the case at hand, no translation is needed, since we already set the first moments to zero. The new coordinates with identity covariance matrix can be found by performing  a Cholesky decomposition $\Theta = L\,L^T$--where $L$ is a lower triangular matrix--of the original covariance matrix. The new set of coordinates are found as $L^{-1}(q,p)^T=(x,y)$. This rotation also affects higher-order cumulants, that must be transformed accordingly. In particular, collecting all fourth moments in a four-tensor $K_{abcd}$, this transforms under rotations as:
\begin{equation}
    K'_{abcd} = \sum_{i,j,k,l} U_{ai} U_{bj} U_{ck} U_{dl} K_{ijkl},
\end{equation}
where $U=L^{-1}$ is the rotation matrix and $K'_{abcd}$ is the rotated tensor and all the indices are binary (either $0$ or 1). In terms of cumulants, the original tensor is defined as:
\begin{equation}
    K_{ijkl} = \kappa_{q^{n_q},p^{n_p}},
\end{equation}
with $n_q$ and $n_p$ being the number of zeros and ones in the indices respectively. 

For long times, the system approaches a steady state, whose moments are fixed and time-independent. Crucially, this asymptotic state retains non-zero higher-order moments, indicating that the system settles into a Non-Equilibrium Steady State (NESS). This state characterizes the long-time behavior of the collapse model, consistent with theoretical expectations. \Cref{fig:Gram_Chalier_Wigner_and_Marginals} shows the time evolution of the Gram-Chalier approximated Wigner function for three different times, taken at $t = 10$, $t = 500$, and $t = 1200$ for $\beta = 4.0$ and considering a Gaussian initial state with $\sigma_p^2(0)=\sigma_q^2(0)=1/2$ and $\sigma_{qp}(0)=0$. The early time behavior ($t = 10$) of the Wigner function closely resembles a Gaussian distribution. The non-Gaussian behavior is more prominent at later times, when the the tails of the distribution show a pronounced discrepancy with Gaussianity.


\subsection{Free particle solution}\label{sec:free particle}
Finding the time evolution governed by \cref{eq: working model} is a prohibitive task, and even finding the stationary solution in the general case turns out to be problematic due to the presence of the fourth order derivative term. However, it is possible to find the analytical solution to the stationary problem at least for the free particle. The free particle is a particularly simple case because, as we will show later in this section, the asymptotic Wigner function is actually independent on $q$. The asymptotic state is thus found by setting to zero, in \cref{eq: working model}, all spatial and time derivatives:
\begin{equation}
\begin{aligned}
&\left(F_{1}+2F_{2}\right)W(p)+\left(F_{1}+4F_{2}\right)p\partial_{p}W(p)+ \\&+(D_{pp}+F_{2}p^{2})\partial_{p}^{2}W(p)=0 \,,
\end{aligned}
\end{equation}

In the limit $F_2\to 0$ this equation reduces to the usual Langevin problem whose steady--state is a Gaussian. However, if $F_2\neq 0$ there is an additional term in the diffusion constant that is multiplicative in the momentum of the particle, $p^2\partial_p^2W(p)$. This term can be regarded as a dynamical or turbulent noise that is responsible for the onset of long--range correlations that translate into a non--equilibrium fat-tailed distribution \cite{schenzle1979multiplicative}. Indeed, by direct substitution, one can check that the solution is proportional to 
\begin{equation}
    W(p)\propto\left(1+\frac{F_2 p^{2}}{D_{pp}}\right)^{-\frac{F_1+2F_2}{2F_2}}
\end{equation}
and it is normalizable provided 
\begin{equation}
    \mathcal{C} \int_{-\infty}^{\infty}\,dp \left(1+\frac{F_2}{D_{pp}}p^2\right)^{-\frac{F_1+2F_2}{2F_2}}<\infty
\end{equation}
which is satisfied if the exponent in the integrand is greater than $1/2$, as is always the case for positives $F1$ and $F2$. Then one easily gets the normalization constant
\begin{equation}
    \mathcal{C} = \sqrt{\frac{F_2}{D_{pp}\pi}} \frac{\Gamma (\frac{F_1+2F_2}{2F_2})}{\Gamma (\frac{F_1+F_2}{2F_2})}\,.
\end{equation}
This type of distribution has been studied in different contexts, such as plasma physics \cite{livadiotis2013understanding,pierrard2010kappa} and high-energy physics \cite{biro2005power}, and it can thus arise from different physical mechanisms that have in common a non--linear character. Requiring this distribution to have a finite second moment, and thus a finite energy, is equivalent to $F_1>F_2$ which, assuming that all other parameters are given, imposes an upper bound on the friction parameter:

\begin{equation}
    \beta < \frac{16 R_0^2m}{9\hbar^2}\,.
\end{equation}

\begin{figure*}
Marginal distribution of the position.\\
    \includegraphics[width=\textwidth]{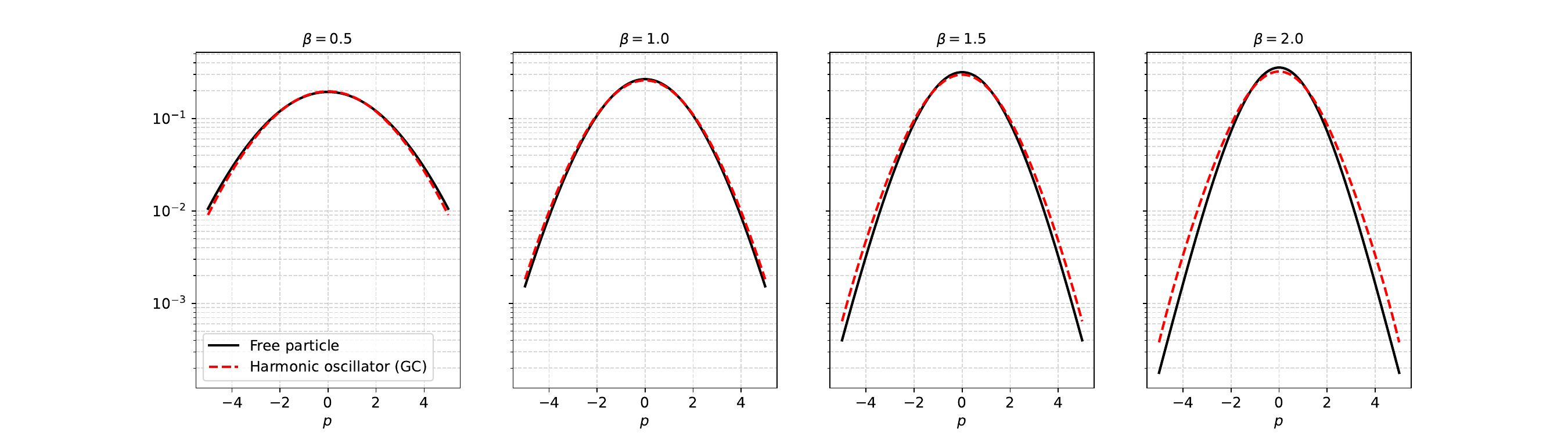}\\
    \caption{Comparison between the steady--state marginal PDF of  the momentum as computed analytically for the free particle (solid line) and numerically for the harmonic oscillator (dashed line). The parameters chosen are $m=\omega = 1.0$, $R_0 = 3$, $\gamma=1$, $\beta = 2$ and $\hbar=k_B=1$ and results are presented in Log scale. Interestingly, the harmonic oscillator and the free particle seem to share a similar marginal over $p$ in the long time limit. It is also possible to see how, as the non--Gaussianty increases with $\beta$, the overlap between the two distributions become smaller.}
    
\label{fig:marginal_analytic}
\end{figure*}

This bound gets tighter if one asks for a finite fourth moment: 

\begin{equation}
    \beta < \frac{4 R_0^2m}{3\hbar^2}\,,
\end{equation}

which is a necessary condition for the applicability our GC approximation in the weak non-Gaussian limit. Note that, of course, the GC expansion could break down for even smaller values of $\beta$. Finally, a numerical analysis shows that this distribution also well approximates the steady--state marginal distribution over $p$ even for a harmonic oscillator, as seen in \cref{fig:marginal_analytic}. In this figure, it is also possible to see how, as the value of $\beta$ gets larger and non--Gaussianity increases, the momentum marginal obtained via the Gram-Charlier expansion of the harmonic oscillator and the momentum distribution of the free particle become less and less similar.

\subsection{Quantifying Non-Gaussianity}
\begin{figure}[ht!]
\centering
\includegraphics[width=1\linewidth]{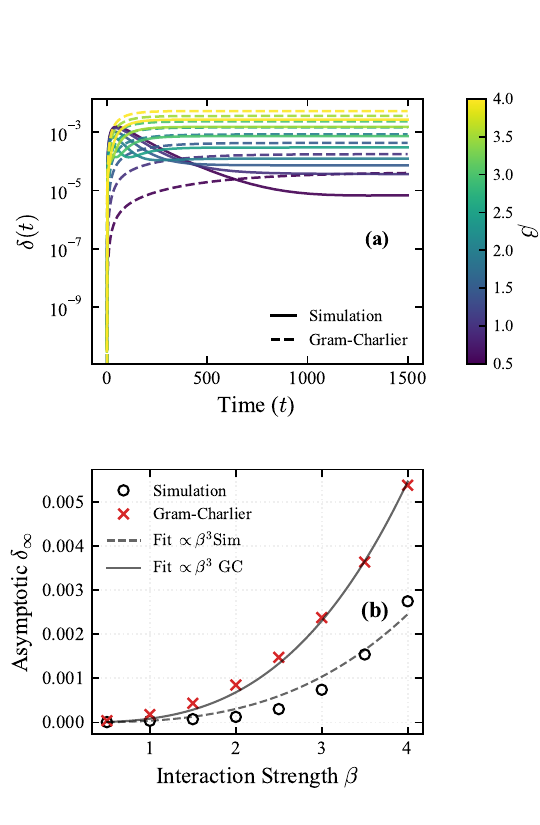}
\caption{Quantitative analysis of the evolution of non-Gaussianity of the states $\delta[\rho]$ for a range of $\beta$ values. The system parameters are set to $m=\omega=1$, $\gamma=1$, $R_0=3$, and $\hbar=k_B=1$, with the system prepared in the ground state at $t=0$. (a) Time evolution of the non-Gaussianity for various dissipation parameters $\beta \in [0.5, 4.0]$. During the transient phase, $\delta[\rho]$ briefly grows larger than its final asymptotic value before decreasing. (b) The asymptotic non-Gaussianity $\delta[\rho_\infty]$ as a function of $\beta$. The GC method actually overestimates the non-Gaussianity of the state according to this measure. A polynomial fit of the data points reveals an exact cubic scaling behavior, $\delta[\rho_\infty] \propto \beta^3$, with a root-mean-square error on the order of $10^{-18}$.}
\label{fig:non_gauss}
\end{figure}

\begin{figure}[ht!]
\centering
\includegraphics[width=1\linewidth]{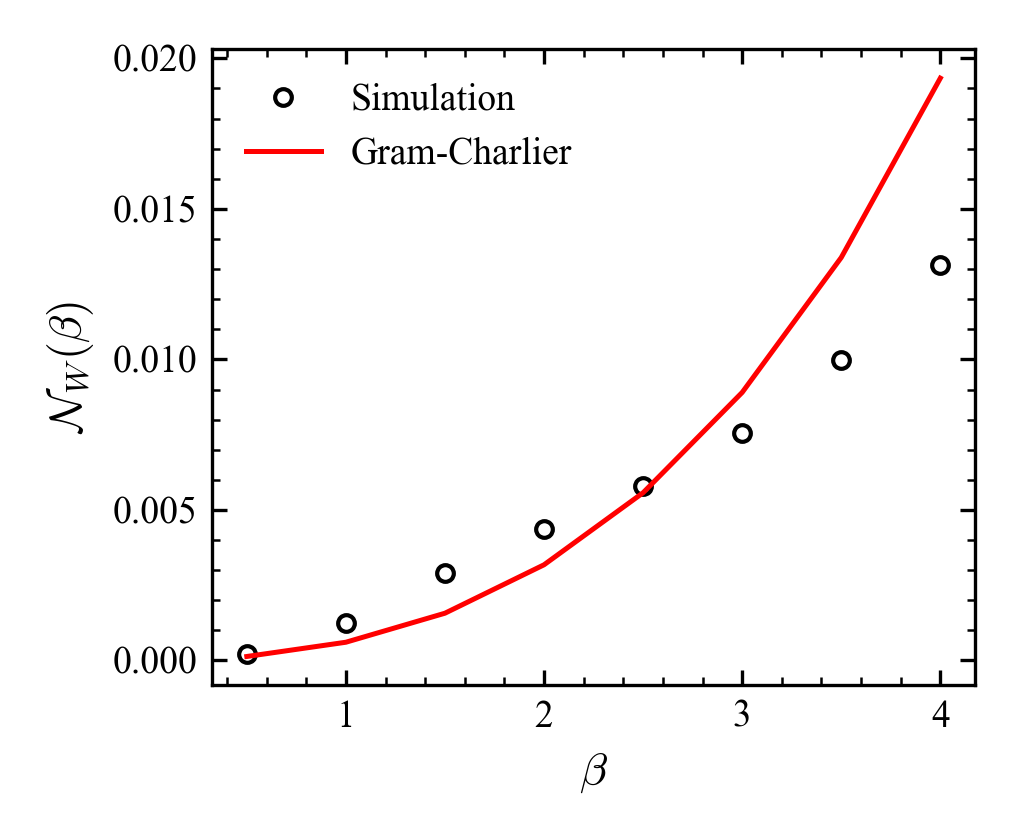}
\caption{Quantitative analysis of the steady--state non-Gaussianity as characterized by the negentropy $\mathcal{N}_W=K(W||W_{ref})$ for a range of $\beta$ values. The system parameters are set to $m=\omega=1$, $\gamma=1$, $R_0=3$, and $\hbar=k_B=1$, with the system prepared in the ground state at $t=0$. With this measure, for small values of $\beta$ the simulated steady--state is more non--Gaussian then the one obtained via GC expansion. For large values of the friction parameter the hierarchy is inverted.}
\label{fig:negentropy}
\end{figure}

In order to quantify the non-Gaussianity of the state throughout the evolution we follow the discussion in \cite{genoni2008quantifying}. There, the non-Gaussianity of a quantum state $\rho$ is defined as

\begin{equation}\label{eq: non gauss meas}
    \delta[\rho] = \frac{D^2_{HS}[\rho,\tau]}{\mu[\rho]}\,,
\end{equation}
where $D_{HS}[\rho,\tau]$ is the Hilbert-Schmidt distance between $\rho$ and $\tau$, the latter being the reference Gaussian state such that the first and the second moments of the two states coincide and $\mu[\rho]=\Tr[\rho^2]$ is the purity. Writing the squared H-S distance as
\begin{equation}
    D^2_{HS}[\rho,\tau]=\frac{\mu[\rho]+\mu[\tau]-2\Tr[\rho\tau]}{2}
\end{equation}

we can compute $\delta[\rho]$ of our state from its approximate Wigner function using the formula for the overlap of states $\Tr[\rho\tau] = \iint \,dq\,dp\, W_{\rho}(q,p)W_{\tau}(q,p)$, where $\tau$ is naturally taken as the reference Gaussian state used for the G-C expansion. The results are reported in \cref{fig:non_gauss} where the same numerical values of \cref{fig:Gram_Chalier_Wigner_and_Marginals} have been used. In \cref{fig:non_gauss}(a) the non-Gaussianity $\delta$ is reported against time for different values of $\beta$ within the range in which the G-C expansion holds. In all these cases, the non-Gaussianity increases during the evolution, finally reaching an asymptotic value. In \cref{fig:non_gauss}(b), the asymptotic non-Gaussianity is reported against $\beta$. A polynomial fit of the data points reveals a cubic behavior $\delta[\rho_{\infty}] \propto \beta^3$. It is worth noting that, despite it being a method to approximate a slightly non--Gaussian distribution, the GC expasion actually overestimate the non--Gaussianity of the state according to this measure. This can be attributed to the effect of higher-order moments that are not captured in our approximations and that are weighted more with respect to the kurtosis by this non--Gaussian quantifier. In particular, even though the tails of the distribution obtained via simulation always decay slower then the ones obtained in the GC approximation, the shape of the distribution around the peak may have a dominant contribution in the computation of $\delta[\rho]$. Considering a different figure of merit for the non--Gaussianity can indeed lead to a different characterization. In \cref{fig:negentropy}, for example, the non--Gaussianity of the steady--state as characterized by the negentropy $\mathcal{N}_W(\beta)$-\textit{i.e.} the Kullback-Leibler divergence between the true distribution and the reference Gaussian one \cite{brillouin2013science}-is reported. Here, the two methods have a crossover in their hierarchy: for small $\beta$ the simulation is characterized by a greater non--Gaussianity, whereas the opposite happens for large values of the friction parameter, the crossover happening between $\beta=2$ and $\beta=3$. We can thus surmise that for small values of $\beta$, $\mathcal{N}_W$ is more susceptible to the the way  the distribution behaves for large deviations then $\delta[\rho]$.

\subsection{Entropy production}
\begin{figure}
    \centering    \includegraphics[width=0.9\linewidth]{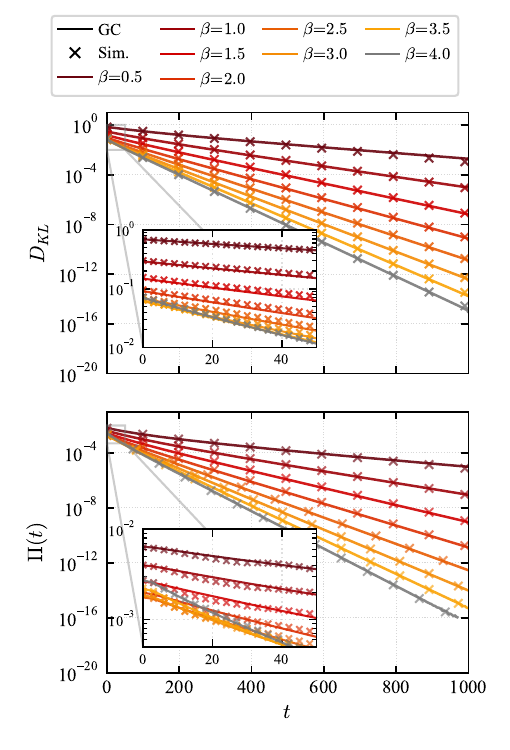}
    \caption{Time evolution of the Kullback-Leibler divergence ($K(W||W_{\infty})$) and the entropy production rate ($\Pi(t)$). The plots display results for dissipation values $\beta \in [0.5, 4.0]$, using parameters $m=\omega=1.0$, $R_0=3$, $\gamma=1$, and $\hbar=k_B=1$. The graphs compare the perturbative Gram-Charlier approximation (solid lines) with the exact pseudo-spectral simulation (cross markers). Both quantities decay with time, and the slope of the decay steepens as $\beta$ increases. In the weak non-Gaussian regime (smaller values of $\beta$), the numerical exact simulation and the Gram-Charlier approximation show good agreement. However, for strong non-Gaussianity ($\beta \ge 3.0$), significant disagreements emerge, as logarithmic dependencies in these thermodynamic quantities heavily penalize inaccuracies in the distribution's tails. At $\beta = 4.0$, the Gram-Charlier approximation ultimately fails to show the correct evolution of the system. The insets highlight the early-time crossing behaviors.} 
    \label{fig:entropy}
\end{figure}

Using \cref{eq: EP}, we can compute the entropy production rate of relaxation dynamics induced by the collapse dynamic in the weakly non-Gaussian regime. In fact, the Wigner function of the state remains positive and thus the entropy is always well defined. The results for different values of $\beta$ are reported in \cref{fig:entropy}, considering an harmonic oscillator of mass $m=1$ and frequency $\omega=1$; the collapse constants are set to $R_0 = 3$, $\gamma = 1$ and the initial state is the ground state of the oscillator. The plots are displayed in logarithmic scale. 
Looking at the plots, it is evident that $\Pi(t)$ remains always positive. Moreover, it looks like the entropy production at a fixed time is always smaller for larger $\beta$. However, this observation turns out to be wrong at a closer inspection. In fact, zooming on early times, we can observe crossings in the various curves for different $\beta$. This signals a change in the initial dissipation regime due to the non--linearity of the relaxation process. Indeed, this feature is not present in standard models of Brownian motion \cite{artini2023characterizing, artini2025nonequilibrium}. 

These results prove that collapse dynamics, in the regime we investigated so far, is consistent with the Second law of Thermodynamics. Interestingly, the steady--state distribution is a highly non--thermal one (see ~\cref{sec:free particle}), suggesting that the system reaches a NESS, and not equilibrium. This in turn implies a constant entropy production \cite{seifert2005entropy,landi2021irreversible}  maintained by the collapse field, acting as an out--of--equilibrium environment. As discussed in \cref{sec: thermo stuff}, this positive, constant entropy production rate needs to be added on top of the entropy production rate induced by the relaxation process, which is computed in this section. Evaluating this extra term, the so called \textit{housekeeping entropy}, would require characterizing the persistent currents in the steady state, which is a particularly involved task in the non--Gaussian setting under consideration. However, being always positive, the housekeeping entropy cannot be the cause any violation of the Second law, thus confirming once more the validity of the model from a thermodynamics standpoint.

\section{Simulation of the non-Gaussian CSL model for strong non-Gaussianity}\label{sec: simulation}

The Gram-Charlier expansion provides a robust approximation for the Wigner function in regimes close to Gaussianity. However, establishing a strict, universal criterion for the failure of this method remains a challenging task. Standard statistical thresholds, such as the Barton-Dennis conditions \cite{Barton1952}, can be used to determine if the truncated polynomial expansion yields a mathematically valid probability distribution based on its moments. Yet, in the context of non-equilibrium thermodynamics, this purely statistical criterion proves insufficient. When calculating high-sensitivity, information-theoretic quantities such as the Wigner entropy production rate and the Kullback-Leibler divergence, significant disagreements emerge between the approximate and exact evolutions at higher dissipation regimes ($\beta \ge 3.0$). Even when the statistical moments fall strictly within acceptable analytical bounds, the logarithmic dependencies in these thermodynamic quantities heavily penalize inaccuracies in the distribution's tails. Consequently, to accurately capture the system's thermodynamic behavior under strong non-Gaussianity, we must move beyond perturbative treatments and solve the full Wigner evolution numerically.

\subsection{Pseudospectral method and exponential time differencing for the Wigner evolution}

The governing partial differential equation \cref{eq: working model} for the Wigner function contains both linear terms with constant coefficients (e.g., standard diffusion) and terms with variable coefficients that depend on the phase-space coordinates, \textit{e.g.}, $p^2 \partial_p^2 W$ or coordinate-dependent drift. To efficiently integrate this equation, we employ a pseudospectral method \cite{boyd2001chebyshev}, which leverages the Fast Fourier Transform (FFT) \cite{Cooley1965, kosloff1986absorbing} for high-accuracy spatial derivatives while evaluating coordinate-dependent multiplications in real space.

Let $W(x,p,t)$ be the Wigner function. The evolution equation can be partitioned as 
\begin{equation}
    \partial_t W = \mathcal{L}W + \mathcal{N}(W),
\end{equation}
where $\mathcal{L}$ represents the linear, constant-coefficient differential operator, and $\mathcal{N}$ contains the coordinate-dependent non-linear functional terms as in \cref{eq: working model}. In Fourier space $(k_x, k_p)$, the spatial derivatives map to simple algebraic products: $\partial_x \to ik_x$ and $\partial_p \to ik_p$.

To perform a computationally lightweight simulation without sacrificing numerical stability, we introduce targeted spectral filtering. A smooth spectral cutoff filter of order 6, given by $\exp[-(R_k/k_c)^6]$, is applied to the unstable cross-derivative terms, where $R_k$ is the normalized radial wavevector. Furthermore, to prevent spectral blocking and aliasing instabilities at the grid scale, an artificial hyperviscosity term proportional to $-k^6$ is added to $\mathcal{L}$ \cite{boyd2001chebyshev}. This acts as an energy sink exclusively at the smallest resolved scales, preserving the macroscopic physical dynamics. An absorbing sponge layer (cosine-squared mask) is also applied near the domain boundaries to prevent unphysical periodic wrap-around artifacts inherent to FFT-based methods \cite{kosloff1986absorbing}.

Standard explicit time-integration schemes, such as Forward Euler, are unsuitable for this system due to severe time-step constraints imposed by the stiff linear diffusive terms. Instead, the time evolution is performed using a fourth-order Exponential Time Differencing Runge-Kutta (ETDRK4) scheme \cite{kassam2005fourth}. This approach solves the linear part exactly while using an RK4 approximation for the variable-coefficient terms. For each time step from $t$ to $t+\Delta t$, the procedure is as follows:

\begin{enumerate}
    \item The linear operator is exponentiated in Fourier space to create the exact propagators $E = \exp(\mathcal{L} \Delta t)$ and $E_{1/2} = \exp(\mathcal{L} \Delta t / 2)$.
    \item To evaluate $\mathcal{N}(W)$, we transform the current state to Fourier space ($W \to \tilde{W}$), compute the necessary derivatives by multiplying by $ik_x$ or $ik_p$, and apply the inverse 2D FFT to return to real space.
    \item Point-wise multiplications with the phase-space coordinate grids (e.g., $x$ or $p^2$) are performed in real space. The resultant Hamiltonian and dissipative terms are summed and transformed back to Fourier space.
    \item The state is advanced by sequentially evaluating the four ETDRK4 sub-steps ($k_1, k_2, k_3, k_4$). The final updated state in Fourier space is assembled by combining these steps with the exact exponential propagators, and then transformed back to real space if required for data exportation.
\end{enumerate}

Despite the use of artificial hyperviscosity and spectral truncation to maintain a lightweight computational grid, the physical fidelity of the simulation is rigorously preserved. To certify the validity of the numerical evolution, the time-dependent statistical moments (e.g., $\langle x^2 \rangle$, $\langle p^2 \rangle$, and $\langle xp \rangle$) are continuously extracted directly from the simulated Wigner distribution and successfully benchmarked against the exact  solutions of the corresponding moment equations of \cref{sec: moments}, as we can see in \cref{fig:moments_Gram_Chalier}.

The numerical solution is computed on a discrete two-dimensional phase-space grid. The grid resolution is $N_x = 320$ points in position and $N_p = 320$ points in momentum, chosen to be powers of two for FFT efficiency. The grid spacing is $\Delta x = 0.4$ and $\Delta p = 0.4$, defining a simulation domain of $x \in [-60.0, 60.0]$ and $p \in [-60.0, 60.0]$. The time evolution is performed with a discrete step of $\Delta t = 2\times10^{-3}$ for a total duration of $t_{\text{final}} = 1500$, resulting in $7.5\times 10^5$ integration steps.

\subsection{Wigner function evolution}

\begin{figure*}
Numerical evolution of the Wigner function \\
    \centering
\includegraphics[width=1\linewidth]{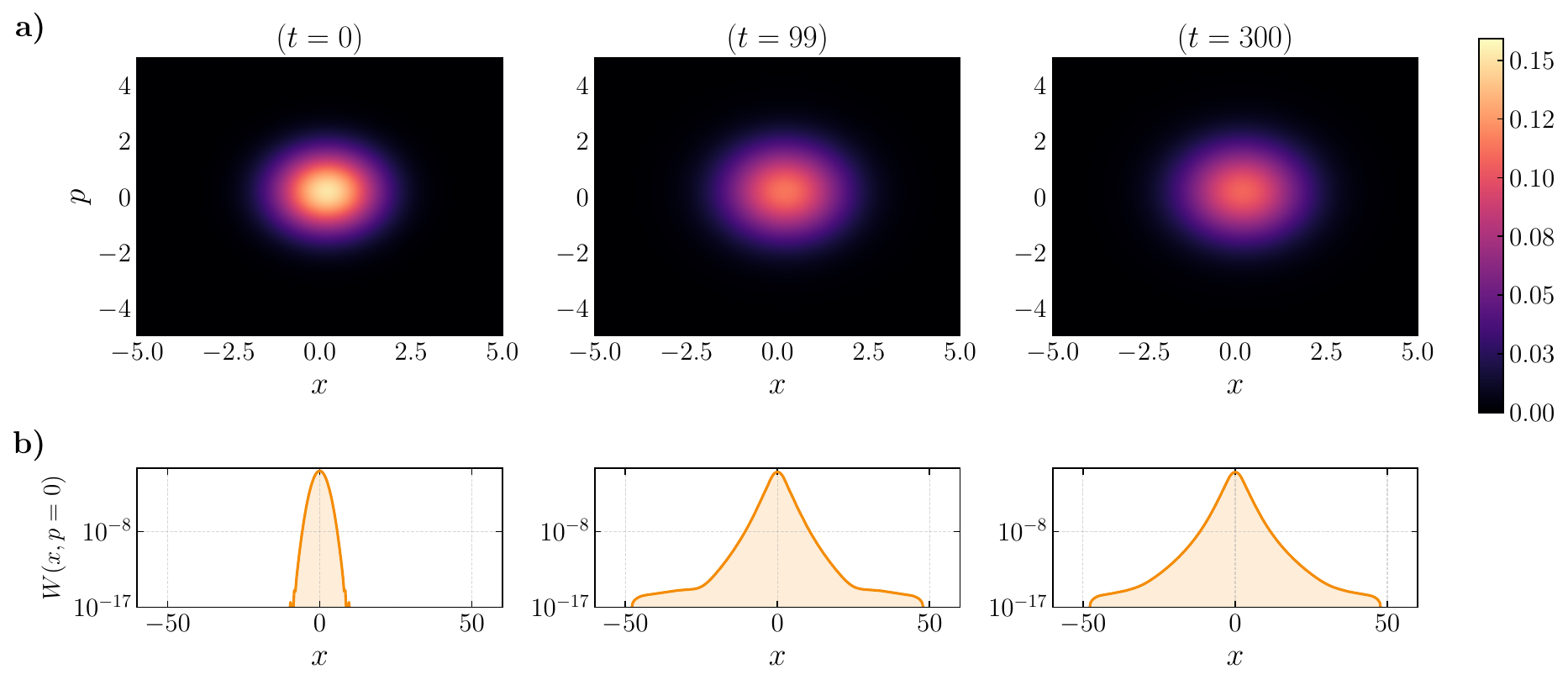}
    \caption{Temporal evolution of the Wigner function for $\beta=4.0$. (a) Phase-space heatmaps at different time steps. (b) Logarithmic slices of the distribution at $p=0$, highlighting the development of non-Gaussian tails.}
    \label{fig:wigner_evolution}
\end{figure*}

We employ the previously described pseudo-spectral scheme to simulate the full phase-space dynamics of the Wigner function $W(x,p,t)$. Starting from an initial Gaussian ground state, the system evolves under the competing influences of diffusion and dissipation. \Cref{fig:wigner_evolution} illustrates the temporal evolution of the distribution for a fixed inverse temperature $\beta = 4.0$.

As shown in \cref{fig:wigner_evolution}(a), the distribution remains centered at the phase-space origin. However, the underlying nonlinear dynamics drives the emergence of significant non-Gaussian features. This departure from Gaussianity is characterized by the formation of heavy tails and a non-zero excess kurtosis, as evidenced in the log-scaled cross-sections presented in~\cref{fig:wigner_evolution}(b). It can be observed that for $t = 99$ and $t = 300$, the slices of the Wigner function already have non-Gaussian tails

This broadening is primarily governed by the higher-order diffusion term, $D_{q^2p^2}$, which acts as a persistent source of tail growth. The numerical results demonstrate that the pseudo-spectral method successfully tracks the evolution of these tails over several orders of magnitude. For the sake of numerical stability, we implement spectral filters that truncate the distribution at a threshold of approximately $10^{-15}$. This cutoff is strategically chosen below the scale of relevant physical features and near the machine precision limit; hence, it prevents numerical divergence caused by the growth of high-frequency components without compromising the essential physics of the model.

By going beyond the limitations of perturbative Gram-Charlier expansions, our exact numerical approach captures the intricate phase-space structures and the precise tail behavior that perturbative methods typically fail to resolve.

\subsection{Kullback-Leibler, entropy production rate, and non-Gaussianity results}

To quantify the thermodynamic implications of the dissipative CSL model and benchmark the limits of our approximation, we analyze the non-Gaussianity measure $\delta[\rho]$, the Kullback-Leibler (KL) divergence $D_{KL}$, and the entropy production rate $\Pi(t)$. 

As shown in Figure 3, the non-Gaussianity dynamically increases before settling into an asymptotic steady-state value. A polynomial fit reveals that this asymptotic non-Gaussianity, $\delta[\rho_\infty]$, scales cubically with the dissipation parameter, such that $\delta[\rho_\infty] \propto \beta^3$. During the transient phase, we also observe an overshooting effect in the exact simulations, where $\delta[\rho]$ briefly exceeds its asymptotic limit before subsequently decreasing.

A comparison between the Gram-Charlier approximation and the exact numerical simulation demonstrates reasonable agreement in the weak dissipation regime ($\beta < 3.0$). Over time, both the Kullback-Leibler distance $D_{KL}$ and the entropy production rate $\Pi(t)$ decay, with the slope of this decay steepening as the value of $\beta$ increases. However, at higher dissipation regimes ($\beta \ge 3.0$), significant disagreements emerge between the analytical approximation and the exact numerical results. For instance, at $\beta = 4.0$, the KL divergence clearly shows that the Gram-Charlier expansion fails to capture the correct temporal evolution of the system. This breakdown occurs because information-theoretic quantities like $D_{KL}$ and Wigner entropy production have logarithmic dependencies that heavily penalize inaccuracies in the non-Gaussian tails of the distribution, even when the statistical moments fall within acceptable analytical bounds.

\section{Discussions and Conclusions\label{sec: conclusions}}

In this work, we investigated the non-equilibrium thermodynamics and phase-space dynamics of the dissipative CSL model. While standard collapse models offer a unified dynamic approach to the quantum measurement problem, they inherently predict an unphysical, indefinite increase in the system's energy. The introduction of a dissipative mechanism successfully resolves this by allowing the system to reach an asymptotic finite energy. However, as we have shown, this linear-friction modification breaks the Gaussian character of the dynamics even in the approximate regime of the Kramers-Moyal expansion, prompting a detailed analysis of the Wigner quasiprobability distribution.

By analytically deriving the moment equations up to the fourth order, we demonstrated that the non-Gaussianity induced by the dCSL model selectively affects the higher-order moments. Specifically, the center-of-mass and the symmetry of the distribution remain entirely unaffected. Instead, the non-Gaussian behavior is driven by the diffusion terms, which act as constant sources that broaden the kurtosis and the tails of the distribution. Consequently, rather than reaching standard thermal equilibrium, the system settles into a Non-Equilibrium Steady State (NESS). We quantitatively characterized this state, revealing that its asymptotic non-Gaussianity, $\delta[\rho_\infty]$, exhibits a strict cubic scaling with the dissipation parameter $\beta$.

To study the thermodynamic evolution, we employed the Wigner entropy framework, allowing us to compute strictly well-behaved, positive entropy production rates. For weak dissipation regimes, a Gram-Charlier expansion proved to be a robust tool to approximate the Wigner function and extract these thermodynamic quantities. However, our analysis reveals a critical limitation of perturbative methods: even when the statistical moments of the distribution mathematically guarantee a valid polynomial expansion, the approximation drastically fails for stronger dissipation regimes (e.g., $\beta \ge 3.0$). Information-theoretic metrics, such as the Kullback-Leibler divergence and Wigner entropy production rate, feature logarithmic dependencies that heavily penalize any inaccuracies in the distribution's non-Gaussian tails.

Therefore, to capture the true thermodynamic behavior of systems subjected to strong collapse-induced dissipation, solving the full partial differential equation using exact numerical methods—such as the pseudo-spectral algorithm implemented here—is strictly necessary. Ultimately, our findings confirm the thermodynamic consistency of the dCSL model—\textit{i.e.} positive entropy production rate $\Pi(t)$—providing a comprehensive framework for tracking its irreversible entropy production and highlighting the crucial role of phase-space tails in quantum macroscopic objectification.

\section*{Acknowledgments.} PBM and SA acknowledge Marco Vetrano for his comments. SA acknowledges Tobias Haas for valuable discussions on the quantification of non--Gaussinanity. PBM acknowledges the support of brazilian agencies CAPES - Finance Code 001, and CNPq, grant No. 140264/2026-4. P.V.P acknowledges the Funda\c{c}$\Tilde{\text{a}}$o de Amparo $\Grave{\text{a}}$ Pesquisa do Estado do Rio de Janeiro (FAPERJ Process SEI-260003/000174/2024). SA acknowledges Tobias Haas for his comments and insights. MP is grateful to the Royal Society Wolfson Fellowship (RSWF/R3/183013), the Department for the Economy of Northern Ireland under the US-Ireland R\&D Partnership Programme, the PNRR PE Italian National Quantum Science and Technology Institute (PE0000023), and the EU Horizon Europe EIC Pathfinder project QuCoM (GA no. 10032223). SD is grateful to the PNRR PE NQSTI (Project QUANTIP PE00000023).
\appendix
\section{Fourth moments coupled ODEs}\label{app:A}

To analyze the evolution of non-Gaussianity, we first compute the evolution of the fourth moments of the Wigner function. Following the derivation of lower moments ODEs of motion in Sec. \ref{sec: moments}, we derive the equations of motion for fourth moments, $\langle q^n p^m \rangle, ~ (n + m =4)$.  The non-Gaussian features manifest at the level of the fourth-order moments, because they are related to the kurtosis of the distribution. The equations are
\begin{align}
    &\frac{d}{dt} \langle q^4 \rangle = 12 D_{qq} \langle q^2 \rangle + \frac{4}{m} \langle q^3 p \rangle,\label{eq: 4 moments 1}\\
    &\frac{d}{dt} \langle p^4 \rangle = 12 D_{pp} \langle p^2 \rangle - 4(F_1 - 3F_2)\langle p^4 \rangle - 4k \langle p^3 q \rangle,\label{eq: 4 moments 2}\\ 
    &\frac{d}{dt} \langle q^2 p^2 \rangle = 2D_{qq}\langle p^2 \rangle + 2D_{pp}\langle q^2 \rangle - 2(F_1-F_2)\langle q^2 p^2 \rangle \nonumber\\
    &\quad + 4 D_{q^2p^2}+ \frac{2}{m}\langle q p^3 \rangle - 2 k \langle q^3 p \rangle,\label{eq: 4 moments 3}\\
    &\frac{d}{dt} \langle p^3 q \rangle = 6 D_{pp} \langle qp \rangle - 3(F_1-2F_2)\langle p^3 q \rangle + \frac{\langle p^4 \rangle}{m} \nonumber\\
    &\quad - 3 k \langle q^2 p^2\rangle,\label{eq: 4 moments 4}\end{align}
    \begin{align}
    &\frac{d}{dt} \langle q^3 p \rangle = 6 D_{qq}\langle qp \rangle - F_1\langle q^3 p \rangle + \frac{3}{m}\langle q^2 p^2 \rangle - k \langle q^4 \rangle.\label{eq: 4 moments 5}
\end{align}
\providecommand{\newblock}{}
\bibliography{refs}

\end{document}